\documentclass[twocolumn,tighten]{aastex631}

\usepackage{amsmath}
\usepackage{xspace}
\usepackage{multirow}
\usepackage{fancyapj}
\usepackage{mathtools}

  {\list{}{\leftmargin=0.1in\rightmargin=0.1in}\item[]}%
  {\endlist}

\newcommand{\E}[1]{\ensuremath{\times 10^{#1}} }

\newcommand{\msol}{\ensuremath{M_{\odot}}\xspace}

\newcommand{\cts}{\rm\,ct\per{s}\xspace}
\newcommand{\kev}{\rm\,keV\xspace}
\newcommand{\hz}{\rm\,Hz\xspace}
\newcommand{\ks}{\rm\,ks\xspace}
\newcommand{\km}{\rm\,km\xspace}

\newcommand{\kpc}{\rm\,kpc\xspace}
\newcommand{\s}{\rm\,s\xspace}

\newcommand{\per}[1]{\rm\,#1\ensuremath{^{-1}}\xspace}
\newcommand{\persq}[1]{\rm\,#1\ensuremath{^{-2}}\xspace}
\newcommand{\lumcgs}{\rm\,erg\per{s}\xspace}
\newcommand{\fluxcgs}{{\rm\,erg{\per{s}}{\persq{cm}}\xspace}}
\newcommand{\fluecgs}{\rm\,erg{\persq{cm}}\xspace}

\newcommand{\nicer}{\textrm{NICER}\xspace}

\newcommand{\src}{IGR~J17062\xspace}
\newcommand{\srcfull}{IGR~J17062--6143\xspace}

\definecolor{deeppurple}{RGB}{153,0,255}

\begin{document}
\nolinenumbers 

\title{On the impact of an intermediate duration X-ray burst on the accretion environment in IGR J17062--6143}

\author{Peter Bult}
\affiliation{Department of Astronomy, University of Maryland,
  College Park, MD 20742, USA}
\affiliation{Astrophysics Science Division, 
  NASA Goddard Space Flight Center, Greenbelt, MD 20771, USA}

\author{Diego Altamirano}
\affiliation{Physics \& Astronomy, University of Southampton, 
  Southampton, Hampshire SO17 1BJ, UK}

\author{Zaven Arzoumanian} 
\affiliation{Astrophysics Science Division, 
  NASA Goddard Space Flight Center, Greenbelt, MD 20771, USA}

\author{David R. Ballantyne}
\affil{Center for Relativistic Astrophysics, School of Physics, 
  Georgia Institute of Technology, 837 State Street, Atlanta, GA 30332-0430, USA}
  
\author{Jerome Chenevez}
\affil{National Space Institute, Technical University of Denmark, 
  Elektrovej 327-328, DK-2800 Lyngby, Denmark}

\author{Andrew C. Fabian}
\affiliation{Institute of Astronomy, University of Cambridge, 
  Madingley Road, Cambridge, CB3 0HA, UK}

\author{Keith C. Gendreau} 
\affiliation{Astrophysics Science Division, 
  NASA Goddard Space Flight Center, Greenbelt, MD 20771, USA}

\author{Jeroen Homan}
\affiliation{Eureka Scientific, Inc., 2452 Delmer Street, 
  Oakland, CA 94602, USA}

\author{Gaurava K. Jaisawal}
\affil{National Space Institute, Technical University of Denmark, 
  Elektrovej 327-328, DK-2800 Lyngby, Denmark}

\author{Christian Malacaria}
\affiliation{Universities Space Research Association, Science and Technology Institute, 
  320 Sparkman Drive, Huntsville, AL 35805, USA}

\author{Jon M. Miller}
\affiliation{Department of Astronomy, University of Michigan,
  1085 South University Avenue, Ann Arbor, MI 48109-1107, USA}

\author{Michael L. Parker}
\affiliation{Institute of Astronomy, University of Cambridge, 
  Madingley Road, Cambridge, CB3 0HA, UK}

\author{Tod E. Strohmayer} 
\affil{Astrophysics Science Division and Joint Space-Science Institute,
  NASA Goddard Space Flight Center, Greenbelt, MD 20771, USA}

\begin{abstract}
We report on a spectroscopic analysis of the X-ray emission from IGR
J$17062-6143$ in the aftermath of its June 2020 intermediate
duration Type I X-ray burst. Using the Neutron Star Interior Composition
Explorer, we started observing the source three hours after the burst was
detected with MAXI/GSC, and monitored the source for the subsequent twelve
days. 
We observed the tail end of the X-ray burst cooling phase, and find that the
X-ray flux is severely depressed relative to its historic value for a three
day period directly following the burst. We interpret this intensity dip as
the inner accretion disk gradually restoring itself after being perturbed by
the burst irradiation. 
Superimposed on this trend we observed a $1.5$ d interval during which the 
X-ray flux is sharply lower than the wider trend. This drop in flux could be isolated to the non-thermal components in the energy spectrum, suggesting that it may be caused by an evolving corona. 
Additionally, we detected a 3.4 keV absorption line at $6.3\sigma$ significance
in a single $472$ s observation while the burst emission was still
bright. We tentatively identify the line as a gravitationally redshifted
absorption line from burning ashes on the stellar surface, possibly
associated with $\prescript{40}{}{\rm Ca}$ or $\prescript{44}{}{\rm Ti}$. 
\end{abstract}

\keywords{%
stars: neutron --
X-rays: binaries --	
X-rays: individual (\srcfull)
}

\section{Introduction}
\label{sec:intro}

Type I X-ray bursts arise from thermonuclear shell flashes in the accreted
envelope of a neutron star \citep{Lewin1993, Strohmayer2006, Galloway2020,
Galloway2021}. It is well established that the observable properties of such
X-ray bursts depend intricately on the characteristics of the accretion, with
the rate, composition, and precise geometry of the accretion flow all playing
a part in shaping the intensity and duration of the X-ray bursts, as well as
their recurrence times \citep{Galloway2021}. Conversely, the intense radiation
field of the X-ray burst can interact with the matter surrounding the neutron
star in various ways, which can significantly perturb the accretion flow
\citep[see][for a review]{Degenaar2018}. Such feedback has received increasing
attention in recent years, as it allows for the bright burst emission to be
used as a probe of accretion processes. For instance, the X-ray burst emission
can reflect off the accretion disk \citep{Ballantyne2004}, change the intensity
and shape of the persistent emission \citep{Zand2013, Worpel2013, Keek2014}, or
perturb the corona causing an intensity suppression at high ($>30\kev$) photon
energies \citep{Maccarone2003, Ji2014, Ji2015}.

Most studies investigating the interactions between the burst
emission and the accretion flow have focused on the brightest
phases of the burst. This focus is natural, as physically one
would expect the largest perturbation to occur when the burst
intensity is highest. A challenge, however, is that the thermal
burst emission often outshines the other radiative processes,
making it difficult to disentangle which spectral changes are
intrinsic to the burst, and which are due to the accretion
flow.  If the accretion environment is sufficiently perturbed,
though, then the timescale on which it relaxes back to its
(preburst) equilibrium state may well be longer than the
cooling timescale of burst emission. Thus, perhaps we can
observe the disk recovery after the bright X-ray burst has
faded. In this paper we present the case of \srcfull (\src), as
an example of what appears to be precisely this scenario.

First discovered in 2007 Jan/Feb \citep{Churazov2007}, \src has
been persistently visible with a low X-ray luminosity of
$\approx6\E{35}\lumcgs$ \citep{Degenaar2017, Eijnden2018}.
Despite its very long outburst history, it was only recently
discovered to be a 163 Hz accreting millisecond pulsar
\citep{Strohmayer2017} in a 38-minute ultra-compact binary
orbit \citep{Strohmayer2018a}. Subsequent monitoring of the
pulsar revealed that the neutron star is steadily spinning-up,
while the binary orbit is rapidly expanding \citep{Bult2021b}.
\src is also one of only a few known systems to show energetic
intermediate duration X-ray bursts \citep{Zand2019}, with three such X-ray
bursts recorded so far. The first burst was detected in $2012$
June and observed near its peak intensity with the BAT and XRT
telescopes aboard the Swift observatory \citep{Degenaar2013}.
These data revealed a significant emission line at $1$\kev, as
well as absorption features in the Fe-K band
\citep{Degenaar2013}, both of which point to the burst emission
reflecting off the accretion disk. Further,
\citet{Degenaar2013} found that the burst light curve showed
a 10-minute episode during which the flux fluctuated by
a factor of 3 over a timescale of seconds. Such variability
episodes are a particularly rare feature of the most energetic
bursts, and may be associated with super-expansion of the
stellar photosphere \citep{Zand2019}. In particular,
\citet{Zand2011} suggested that such variability episodes may
be caused by cloud-like structures above the disk, which
intermittently scatter the burst radiation into or out of the
observer line of sight. 

A second burst was detected in 2015 \citep{AtelNegoro15, AtelIwakiri15} and studied
in detail by \citet{Keek2017}. These authors estimated the source flux during
its photospheric radius expansion (PRE) phase and derived a source distance of $7.3\pm0.5$ kpc.
From the burst fluence, they further estimated the burst ignition column to
be $\approx5\E{10}$g\persq{cm}, indicating the bursts are powered by the ignition of
a thick helium layer deep in the stellar envelope. Finally, they found that at the end of
the burst the flux dropped below the cooling trend before returning to the
long-term persistent luminosity after about four days,
suggesting a disruption of the accretion flow that outlasts the duration of the
X-ray burst itself. 

On 2020 June 22, MAXI/GSC detected a third X-ray burst from \src
\citep{AtelNishida20a}, offering a new opportunity to investigate the impact of
these powerful X-ray bursts on the accretion environment. We therefore
executed a follow-up monitoring campaign with the Neutron Star Interior
Composition Explorer (\nicer; \citealt{Gendreau2017}). Our observations began
about $3$ hours after the MAXI trigger, and continued to follow the source
evolution for $12$ days. A timing analysis of the pulsar properties during
this epoch was previously included in \citet{Bult2021b}. In this paper we
present spectroscopic analysis of these data.

\section{Observations and Data Processing}
\label{sec:data}

We observed \src with \nicer between 2020 June 22 and 2020 July 7 for a total
unfiltered exposure of $40\ks$.  These data are available under ObsIDs
$30341001nn$, where $nn$ runs from $01$ through $12$, with each
ObsID storing all continuous pointings collected over the course of one day.
Throughout this paper we will refer to these ObsIDs using just the final two
digits. All data were processed using \textsc{nicerdas} version 7a, as released
with \textsc{heasoft} version $6.27.2$, using the most recent version of the
instrument calibration (release $20200727$). Following the standard screening
criteria, we filtered the data to retain only those epochs collected when the
pointing offset was $<54\arcsec$, the bright Earth limb angle was $>30\arcdeg$,
the dark Earth limb angle was $>15\arcdeg$, the rate of reset triggers
(undershoots) was $<200$\,ct\per{s}\per{det}, and the instrument was outside of
the South Atlantic Anomaly. By default, the pipeline also attempts to reduce
the background contamination by filtering on the rate of high energy events
(overshoots). In the case of \src this overshoot filtering was found to be too
conservative, introducing many spurious 1-s gaps into the data (see also
\citealt{Bult2021b}).  We therefore applied a more relaxed screening approach
in which we first smoothed the overshoot rate using 5-second bins to reduce
noise, and then retained only those epochs when the absolute overshoot rate was
$<1.5$\,ct\per{s}\per{det} (default threshold is $1.0$), and $<2 \times
\textsc{cor\_sax}^{-0.633}$\,ct\per{s}\per{det} (default scale is 
$1.52$)\footnote{The \textsc{cor\_sax} parameter is
a measure for the cut-off rigidity of Earth's magnetic field in units of GeV\per{c}.}.
Using these screening criteria, we were left with $26\ks$ of good time
exposure. 

\section{Light curve}
\begin{figure}[t]
  \centering
  \includegraphics[width=\linewidth]{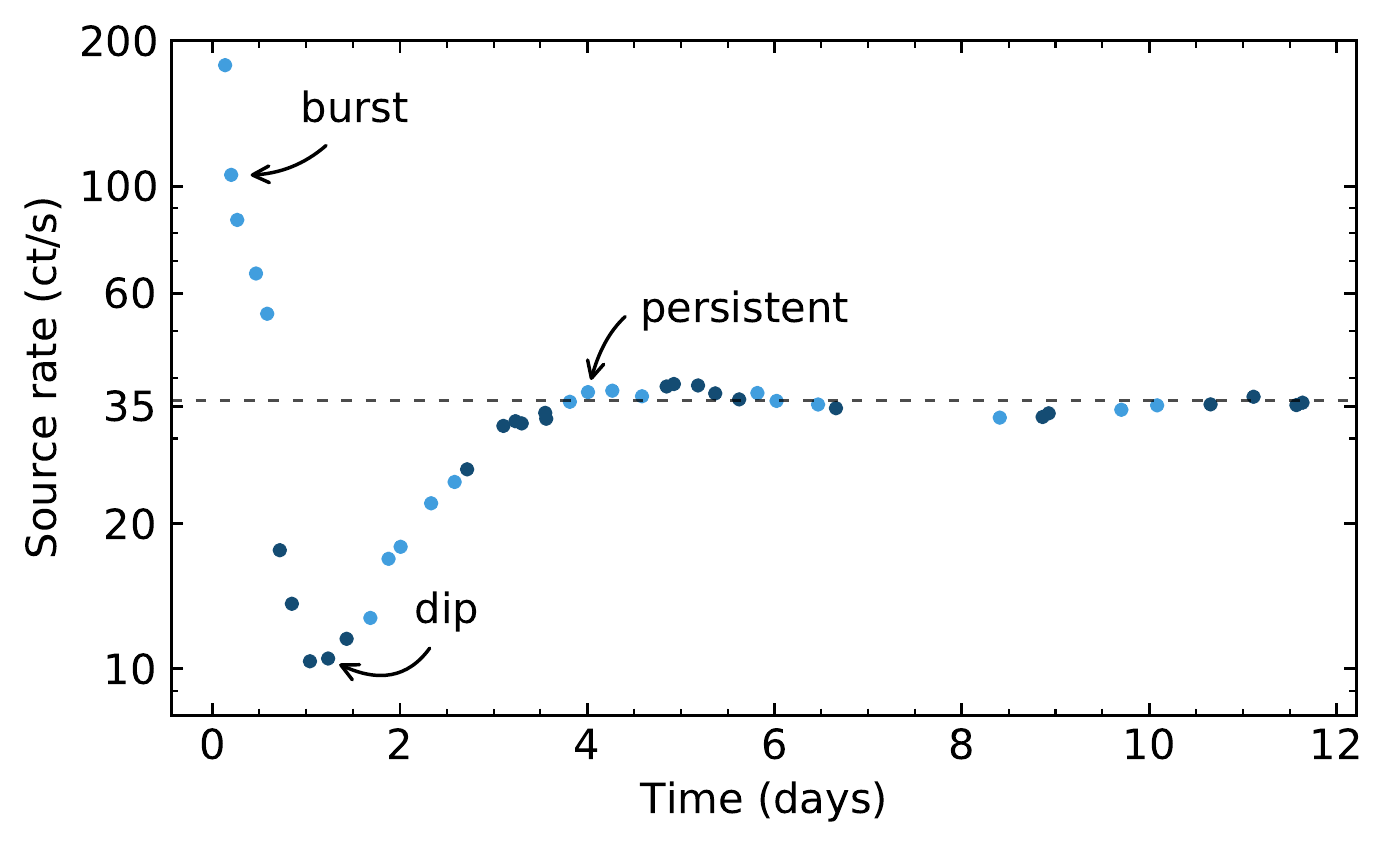}
  \caption{%
    Background subtracted $0.3-6\kev$ \nicer light curve of \src relative to
    the MAXI/GSC trigger on 2020 June 22 (MJD 59022.34403,
    \citealt{AtelNishida20a}). Each point shows the average count-rate of
    a single continuous pointing, while the alternating colors indicate the
    even and odd numbered ObsIDs. The black dashed line shows the count rate
    observed in 2020 August \citep{Bult2021b}. The energy spectra of the three  
    highlighted pointings are shown in Figure \ref{fig:example spectra}.
  } 
  \label{fig:lc}
\end{figure}

To construct a light curve of our \nicer observations, we grouped the data by
continuous pointing. Across the 12 ObsIDs included in this analysis, there were
$43$ such pointings. The good time exposure per pointing ranges between
$50-1200\s$, with the majority of exposures clustered around $400\s$ and
$900\s$. For each pointing we proceeded to extract an energy spectrum and
generated an associated background spectrum using the 3C50 background model
(Remillard et al., 2021, submitted)\footnote{
  \url{https://heasarc.gsfc.nasa.gov/docs/nicer/tools/nicer_bkg_est_tools.html}
}. The source is detected above the background level between about $0.3-6.0\kev$, 
hence, we calculated the background subtracted source rate in this energy band. In 
Figure \ref{fig:lc} we show the resulting light curve, where we expressed all observation 
times relative to the MAXI burst trigger. 

Further observations of \src were collected with \nicer over a two week period
in 2020 August, at which time the mean count rate was found to be $36\pm1\cts$
\citep{Bult2021b}. We adopt this rate as our estimate of the long-term
persistent (non-burst) rate. Comparing the present observations with this
persistent rate, we see from Figure \ref{fig:lc} that the source was initially
detected well above the long-term mean intensity, but showed a rapid decay in
its count rate. At $t\approx0.6{\,\rm days}$, this decay evolves into a sharp
drop: during a $3$ hours data gap between pointings, the source rate decreased
from $55\cts$ to $18\cts$ -- well below the persistent rate. Over the following
three-day period, we observed the source reach a minimum count rate of about
$10\cts$ before recovering back to the persistent rate of $36\cts$. 
From $t=4\,{\rm d}$ and onward, the observed count rate appears to show a small
amplitude oscillation around the long-term mean, however, we caution against
over interpreting this trend; it is entirely consistent with gradual intensity
variations seen in long-term \nicer monitoring of this source \citep{Bult2021b}.

Finally, we constructed a source light curve using a 1-s time resolution and
searched for periods of rapid variability similar to the episode observed by
\citet{Degenaar2013}. No such variability features were found.
We note, however, that the observations of \citet{Degenaar2013}
were collected minutes after the burst ignition, whereas the \nicer observations
presented here were collected hours later. Hence, we did not sample the same
phase of the X-ray burst.

\section{Spectroscopy}
  Turning our attention to the spectral properties of \src, we considered the
  energy spectra extracted from each pointing and modeled them using
  \textsc{xspec} version v12.11 \citep{Arnaud1996}. In this analysis we described
  the interstellar absorption using the T\"ubingen-Boulder model
  (\texttt{tbabs}; \citealt{Wilms2000}), fixing the absorption column density
  to $N_H = 1.1\E{21}\persq{cm}$ \citep{Bult2021b}.

\begin{figure}[t]
    \centering
    \includegraphics[width=\linewidth]{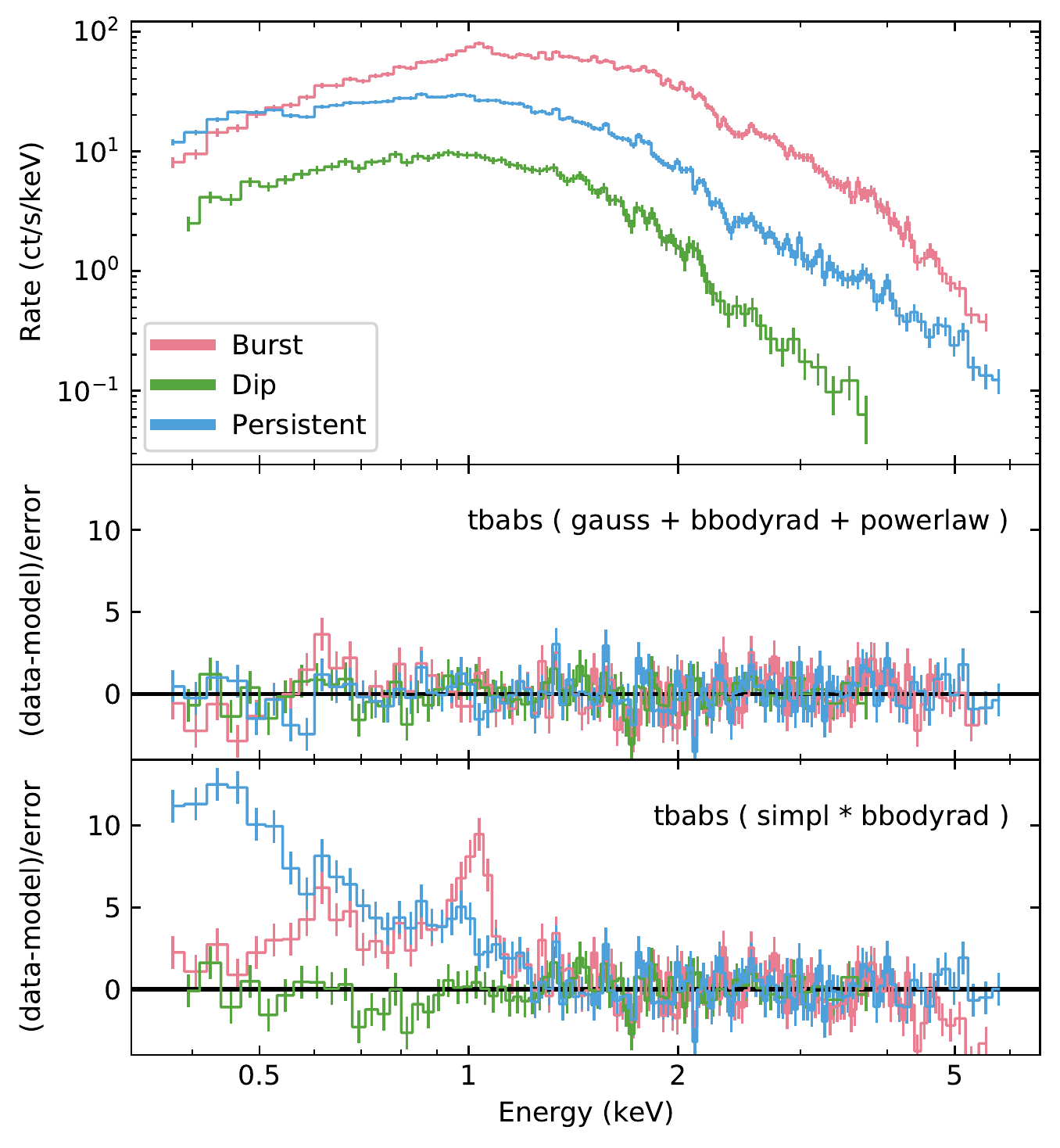}
    \caption{%
      Example spectra of \src at three stages in the light curve evolution.
      Specifically, we show the spectrum of the burst cooling tail (red; ObsID
      01, pointing 2, $412\s$ exposure), the spectrum at the bottom of the
      intensity dip (green; ObsID 02, pointing 4, $828\s$ exposure), and the
      spectrum obtained once the source has recovered to its persistent state
      (blue; ObsID 05, pointing 2, $960\s$ exposure). In the top panel we show
      the energy spectrum, while the middle and bottom panel show the
      error-weighted residuals for the spectral models listed. For the bottom
      panel we fit the model to photon energies $\gtrsim1.7\kev$ to illustrate
      the magnitude of the soft energy flux excess (see text for more details). 
    } 
    \label{fig:example spectra}
  \end{figure}

  For illustrative purposes, we first show an example spectrum from three
  distinct phases of the light curve evolution in Figure \ref{fig:example spectra}.
  Specifically, we show: the burst cooling tail (ObsID 01, pointing 2), the
  minimum of the intensity dip (ObsID 02, pointing 4), and the persistent
  spectrum after the intensity has recovered (ObsID 05, pointing 2). From
  this figure it should be clear that the spectral shape evolves through time,
  with the persistent spectrum being especially more pronounced at the lowest
  and highest photon energies. 

  Prior spectroscopic studies of \src have consistently found that the X-ray
  continuum can be well described with a phenomenological model consisting
  of an absorbed blackbody plus power law \citep{Degenaar2013, Degenaar2017,
  Keek2017, Eijnden2018, Bult2021b}. The burst spectra are further found to exhibit an 
  emission line observed at $1\kev$ \citep{Degenaar2013,Keek2017}, while the 
  spectra of the persistent emission typically show the
  same $1\kev$ emission line as well as a $6.5\kev$ emission line
  \citep{Degenaar2017,Eijnden2018}.  In Section \ref{sec:powerlaw fits} we
  apply this model to our data. 

  A more complex, but physically appropriate description of the X-ray emission
  from \src characterizes the spectrum using a blackbody plus Comptonization and
  a disk reflection component \citep{Degenaar2017,Keek2017,Eijnden2018}. In
  Section \ref{sec:reflection fits}, we therefore model the spectra without the
  power law component, and instead convolve the blackbody component with 
  the \texttt{simpl} Comptonization model \citep{Steiner2009}. We further
  add a disk reflection component, the specifics of which we discuss in more
  detail below. 

\subsection{Phenomenology}
\label{sec:powerlaw fits}
\begin{figure}[t]
  \centering
  \includegraphics[width=\linewidth]{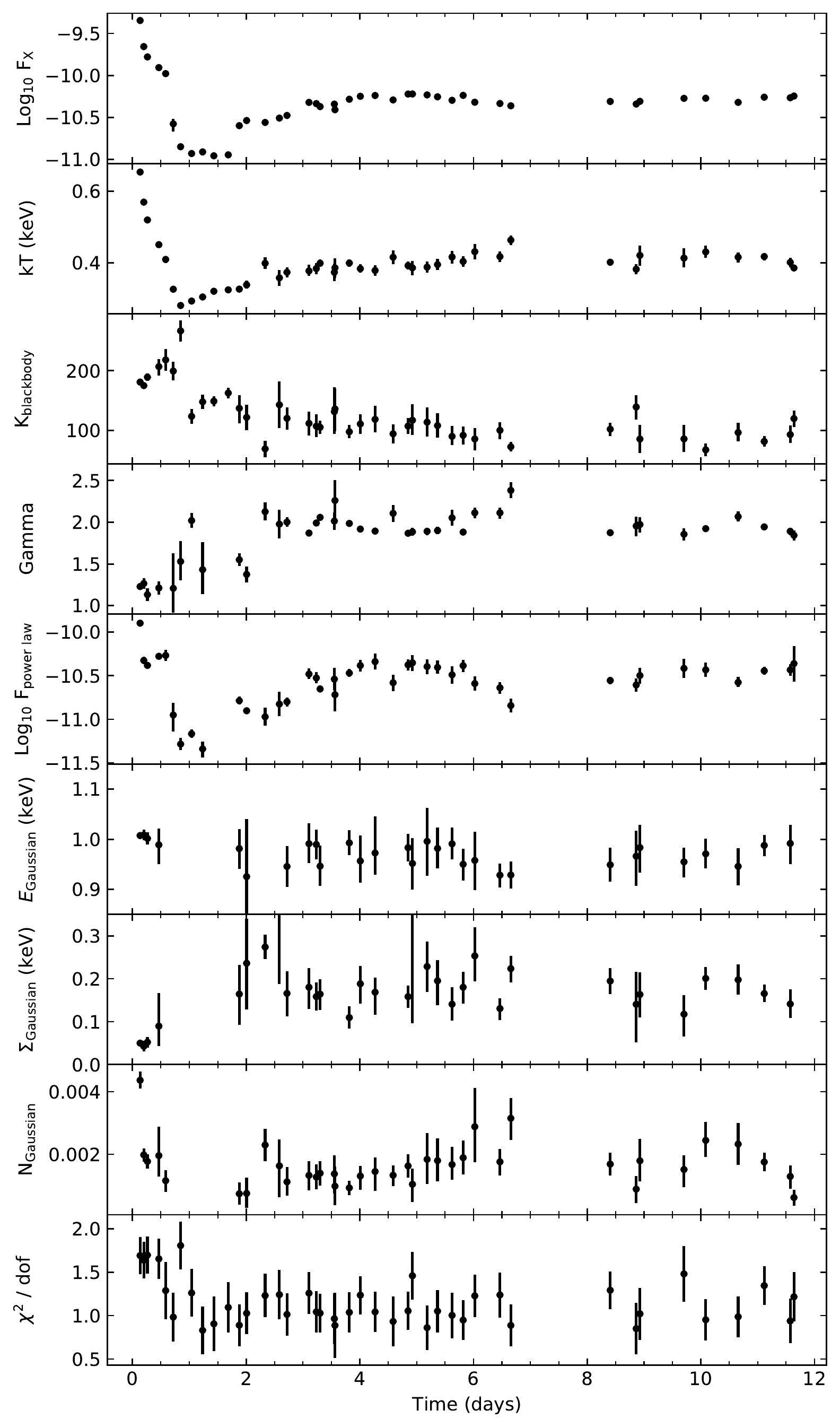}
  \caption{%
    Spectral parameter evolution of \src in terms of the phenomenological model
    (see Section \ref{sec:powerlaw fits}). We show, from top to bottom: the
    $1-10\kev$ X-ray flux, the blackbody temperature, the blackbody
    normalization (in units of $(\km/10\kpc)^2$), the power law photon index,
    the $1-10\kev$ power law flux, the Gaussian line energy, width,
    normalization (in units of ct\per{s}\persq{cm}), and finally, the fit
    statistic. All error bars show $1\sigma$ uncertainties. For the fit
    statistic this uncertainty is calculated from the width of the $\chi^2$
    distribution as $\sigma = \sqrt{2/{\rm dof}}$.
  } 
  \label{fig:pl spectra}
\end{figure}

  We modeled each of the \nicer spectra in the $0.3-6.0\kev$ range using the
  absorbed blackbody plus power law model, with an added Gaussian to model
  the $1\kev$ emission line. This model generally does a very good job
  at describing the spectra (see middle panel of Figure \ref{fig:example
  spectra}), although the burst dominated data show systematically large 
  reduced $\chi^2$ (see Figure \ref{fig:pl spectra}, bottom panel). The first
  pointing further shows more complex line features, which we investigate
  separately in section \ref{sec:lines}. Finally, we find that the Gaussian
  component is not always required to obtain a good fit. 
 
  As a formal test for whether or not the Gaussian emission line should
  be included, we first fit a spectrum without the line. We then generate
  500 simulated spectra from this model and calculate the $\Delta\chi^2$
  improvement obtained after adding a $1\kev$ line to the spectral model.
  Similarly, we calculate the $\Delta\chi^2$ improvement obtained from
  the real data. If the $\Delta\chi^2$ of the real data measurement is 
  larger than $95\%$ the simulated realizations, we include the line
  in our model, and otherwise we leave it out. 
  Using this formalism we find that the $1\kev$ line is present in all spectra
  except for the five pointings of ObsID 02 and the first pointing of ObsID 03. 

  In Figure \ref{fig:pl spectra} we show the best-fit spectral parameters as
  a function of time along with the $\chi^2$ fit statistic obtained for each
  fit. Example residuals are shown in the middle panel of Figure
  \ref{fig:example spectra}.
  Based on these results, we can divide the evolution of \src into three
  stages: a high flux stage below $t=0.6$\,d; a very low flux stage between
  $t=0.6-1.9$\,d; and a return to the persistent spectral shape after
  $t=1.9$\,d. Of all spectral parameters, only the blackbody temperature shows
  a smooth evolution across all three stages: it decays from $0.65\kev$ down to
  a minimum of $0.3\kev$, before gradually climbing back to its persistent
  value of $0.4\kev$. 
  All other parameters are distinctly different at early and late times.
  Initially the spectrum is dominated by the thermal burst emission, while showing
  a low power law photon index and a narrow $1\kev$ emission line. During
  the low flux stage the intensity of the power law drops and softens, while the emission line disappears.
  Finally, after exiting the low-flux stage, the power law becomes more prominent, the
  $1\kev$ emission returns, and the overall spectral shape recovers its persistent
  morphology.

  The question arises if the temporary non-detection of the $1\kev$ line is due
  to limited sensitivity associated with the lower count rates or if it
  indicates a physical weakening of the line. To investigate this question, we
  consider the $95\%$ upper limits on the line normalization, finding
  them to be comparable to the $1\sim2\E{-3}\,{\rm ct}\per{s}\persq{cm}$ measured
  when the line is significantly detected. Additionally, we stacked all data
  in which the line was not directly observed, which yielded a single energy
  spectrum with $3.7$\ks exposure. This stacked spectrum still did not show an
  emission line at $1\kev$ line, to a 95\% upper limit on the line
  normalization of $2\E{-4}\,{\rm ct}\per{s}\persq{cm}$. Our upper limit is
  about $5-10$ times below the typical line strength in persistent emission,
  hence, the $1\kev$ emission line indeed weakens as the source moves through
  its low intensity phase.

\subsection{Disk reflection}
\label{sec:reflection fits}

  For our reflection component,
  we adopt the photoionized reflection model of \citet{Keek2017}. This model is
  based on the earlier work of \citet{Ballantyne2004} and describes the
  reflected emission from a dense helium-rich accretion disk that is illuminated
  by a blackbody of variable temperature. Specifically, the model assumes
  a high number density of $n = 10^{20} {\rm cm}^{-3}$, and covers blackbody
  temperatures between $0.2-1.2\kev$. The disk ionization parameter ranges
  between $\log \xi = 1.5-3.0$, where $\xi = 4 \pi F / n$ with $F$ the flux
  of the illuminating blackbody continuum and is expressed in
  units of erg\per{s}cm. We will refer to this model component as
  \texttt{bbrefl}. To account for relativistic broadening, we further
  convolve the reflection component with the \texttt{relconv} convolution
  model \citep{Dauser2013}. This model includes a number of parameters
  describing the viewing geometry of the accretion disk, which are important for
  shaping any relativistically broadened emission lines included in the spectral
  model. Because we do not observe an Fe K$\alpha$ line, and the $1.0\kev$ line
  is not accounted for in the reflection model, the model parameters are all
  poorly constrained. Instead of fitting for these parameters, we simply fix
  them to reasonable values. Specifically, we fix the emissivity index at $3$,
  and set the spin parameter to $0.075$, the latter being derived from the 
  known $164\hz$ spin frequency and assuming a $1.4\msol$ stellar mass
  and $10\km$ stellar radius. Additionally, we set the disk inner radius and
  inclination at $50\km$ and $30\arcdeg$, respectively \citep{Bult2021b}. 
  For the assumed stellar mass and radius, this
  inner disk radius converts to $25$ gravitional radii ($r_g = GM/c^2$,
  with $M$ the stellar mass). The disk outer radius is held fixed 
  at the default value of $400r_g$.
  
  We first estimate the blackbody and Comptonization continuum by fitting the
  \texttt{simpl} model to the $>1.5\kev$ data. If we extrapolate this model to
  energies below $1.5\kev$, we typically see an excess of soft emission,
  although the magnitude of that excess varies at the different stages of the
  light curve (see bottom panel of Figure \ref{fig:example spectra}). To
  account for this excess, we add the relativistically broadened
  \texttt{bbrefl} component with its temperature parameter tied to the 
  continuum blackbody, along with a $1\kev$ Gaussian as needed. Refitting
  this model to the whole energy range, we find that the reflection component
  is significantly detected in the first five pointings, but in all cases the
  disk ionization is poorly constrained and pegs at $\log \xi = 3$, the upper
  limit of the parameter range. Once the source transitions into the intensity
  dip, the reflection is no longer required, and the data is adequately
  described by just the Comptonized blackbody continuum. When the source
  intensity is rising again (around $t=1.9$ d), the excess flux at low
  photon energies returns and the continuum alone can no longer satisfactorily
  fit the data. Interestingly, though, the \texttt{bbrefl} component is unable
  to describe the soft excess seen in any of the $t>2$ d spectra, consistently
  yielding reduced $\chi^2$ scores of $2\sim3$ or higher. 

\begin{figure}[t]
  \centering
  \includegraphics[width=\linewidth]{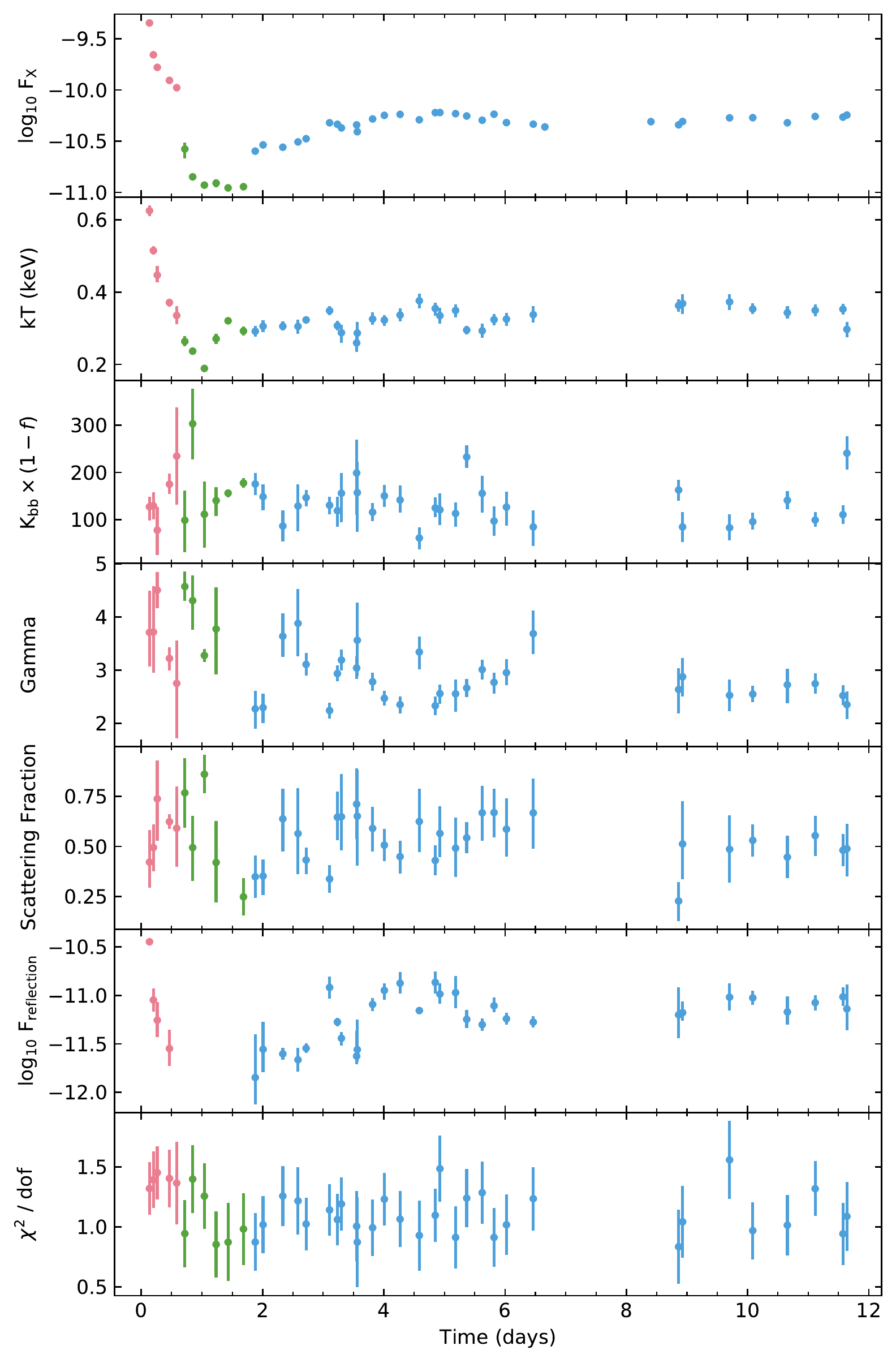}
  \caption{%
    Spectral parameter evolution of \src in terms of the disk
    reflection model (see Section \ref{sec:reflection fits}).
    We show, from top to bottom: the $1-10\kev$ X-ray flux, the
    blackbody temperature, the unscattered fraction of the
    blackbody normalization (in units of $(\km/10 \kpc)^2$), the
    photon index of the Comptonization, the Compton
    scattering fraction, the $1-10\kev$ flux in the disk reflection component,
    and the fit statistic. The color coding indicates the reflection model in
    use, with red for the blackbody continuum reflection, green for no
    reflection, and blue for power-law continuum reflection. All error bars
    show $1\sigma$ uncertainties (see also Figure \ref{fig:pl spectra}).
  } 
  \label{fig:reflection spectra}
\end{figure}

  A possible reason that the reflection model works well for the early data and
  fails for the later data is that the continuum emission illuminating the
  accretion disk is evolving. Some evidence for this can be found in the
  phenomenological fits, which show that the early data is dominated by the
  blackbody, while the later data has a strong power law contribution. In a
  second approach to the reflection modeling, we therefore replace the \texttt{bbrefl} 
  component with \texttt{relxillD} \citep{Dauser2016, Garcia2016} - a reflection
  model for a dense disk that assumes a power law continuum for the incident
  emission. As before, we keep the parameters associated with relativistic broadening
  fixed, and further hold the disk Fe abundance at unity, while setting the
  number density to its highest value of $n = 10^{19} {\rm cm}^{-3}$.
  Following the same procedure as before, we first fitted the continuum to the
  high energy data, then we added the reflection component (and a $1\kev$
  Gaussian as needed), and finally fitted the model to the whole energy range.
  Applying this approach to all spectra, we find the opposite result as before:
  the power law reflection model performs poorly for the soft
  excess seen in the earliest spectra, but gives an excellent description for
  all spectra at $t>1.9$ d.

  In Figure \ref{fig:reflection spectra} we show the spectral parameters
  obtained from the reflection modeling. We combine the \texttt{bbrefl} and
  \texttt{relxillD} models by showing the parameters of the best-fit model for 
  each spectrum. Hence, the early data shows the blackbody reflection (red),
  the low flux points do not include either reflection model (green) and the
  late data shows the power law reflection model (blue).

\subsection{Spectral lines}
\label{sec:lines}
  During the first $472\s$ pointing of our data set, \src was brightest
  and showed a number of narrow line features that were not observed at
  later times. Because of its higher intensity, the source could be
  significantly detected above the background level between $0.25-9.0\kev$,
  hence we consider the spectrum in this slightly wider passband. We initially
  fit the spectrum with the Comptonization model (\texttt{simpl} times
  blackbody), and then ignored all energy channels where the data showed large
  residuals with respect to the model, i.e. $<1.5\kev$, $3-4\kev$.
  Additionally, we mask the Fe-K region ($5.5-7.5\kev$) in case iron
  line emission is causing a bias in the power law tail. 

\begin{figure}[t]
    \centering
    \includegraphics[width=\linewidth]{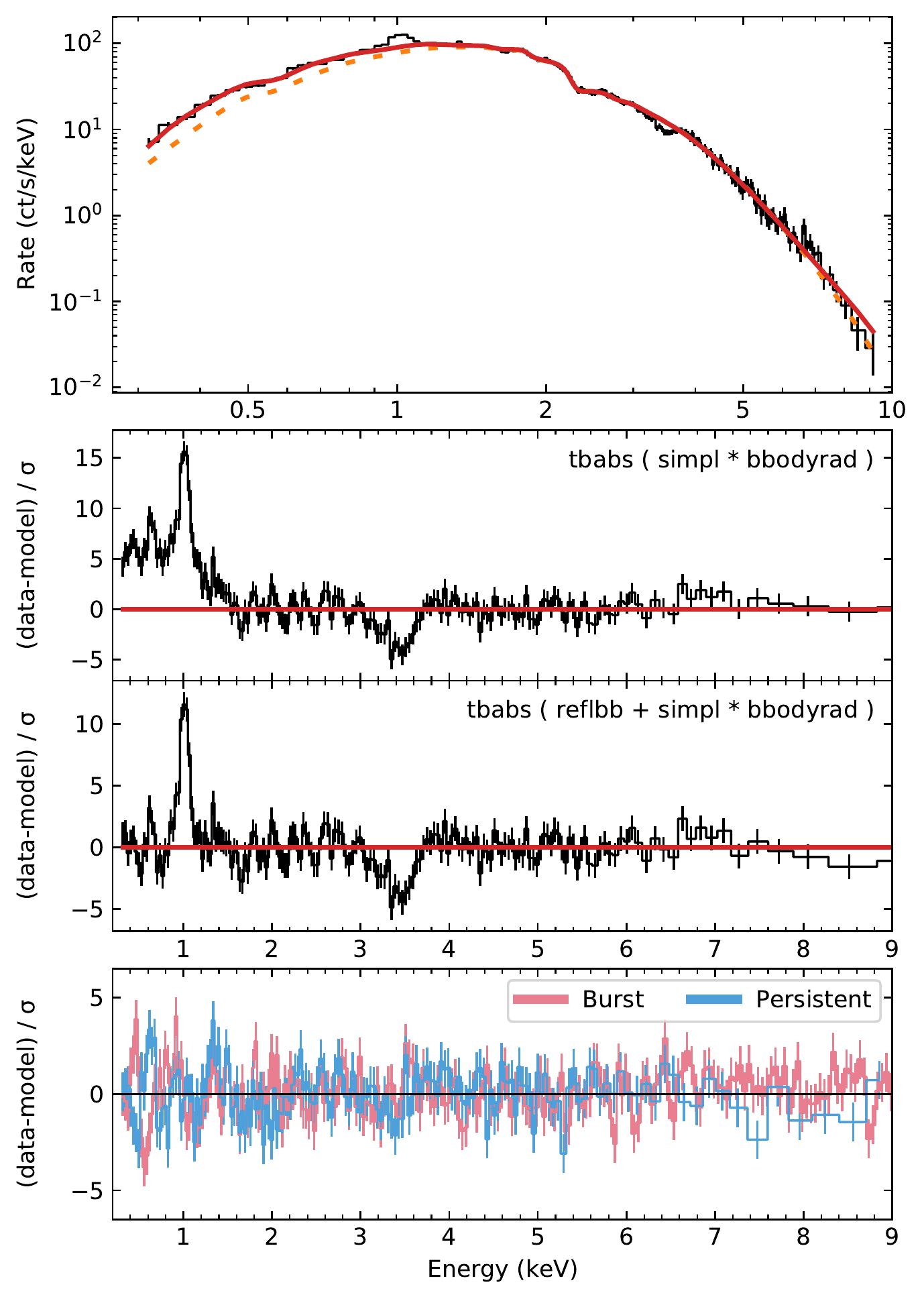}
    \caption{%
      A study of spectral lines in ObsID 01 orbit 1 ($472\s$ exposure). The top
      panel shows the energy spectrum along with the continuum model excluding
      (red) and including (dashed orange) disk reflection. The two middle
      panels show the error-weighted residuals for these respective models on
      a linear scale. For reference, the bottom panel shows the best-fit
      residuals of the stacked spectra (red: Obsid 01 and 02, excluding the
      first orbit; blue: ObsID 05-12), in which the absorption line is not
      evident.
    } 
    \label{fig:spectral lines}
  \end{figure}

  Including the previously ignored energy channels, we observe prominent spectral
  lines relative to the continuum. At low energies, we can observe broad
  soft excess flux and spectral emission features at $0.6\kev$ and $1.0\kev$.
  An absorption feature is resolved at $3.4\kev$, and a weak excess can be
  seen around $6.9\kev$  (see Figure \ref{fig:spectral lines}, top and middle panels). None of these line energies coincide with instrumental features.
  
  To account for the broad low energy excess, we mask the energy channels
  around the spectral lines ($0.5-0.7\kev$, $0.9-1.1\kev$, $3-4\kev$, $5.5-7.5\kev$) and add the
  relativistically broadened \texttt{bbrefl} reflection component to the model.
  After fitting this model, we again included the line regions, finding that
  the reflection model cannot account for any of the observed lines (see
  Figure \ref{fig:spectral lines}, top and bottom panels). 

  Of the four line features observed in the spectrum, the $1\kev$ emission
  line is the most prominent. Adding a Gaussian line component to describe
  this feature improves the fit statistic by $\Delta\chi^2 = 523$, which by
  far exceeds the $\Delta\chi^2$ distribution obtained from a sample of 
  $500$ simulated spectra (see Section \ref{sec:powerlaw fits} for details).
  Additionally, we calculate the simple significance as the line normalization 
  over its $1\sigma$ uncertainty, finding $15.2\sigma$. The best-fit parameters
  of the Gaussian line are reported in Table \ref{tab:spectral lines}.
  Considering the residuals shown in Figure \ref{fig:spectral lines}, we
  note that the profile of the $1\kev$ line is noticeably asymmetric. We note that both the line significance and asymmetry
  are independent of the choice of continuum. Repeating the above analysis using
  the simpler blackbody plus power-law model yields the same results. 

\begin{table}[t]
    \centering
    \caption{%
      Gaussian line parameters
	\label{tab:spectral lines}
    }
    \begin{tabular}{l l l l}
      \hline \hline
	  Line Energy & Width & norm      & Eq. Width  \\
	  (keV)       & (keV) & ($\times 10^{-3}$)  & (keV)      \\
      \tableline
	  $0.630 \pm 0.005$ & $<0.03$         & $0.9 \pm 0.2$ & $0.012 \pm 0.004 $ \\
	  $1.006 \pm 0.003$ & $0.039 \pm 0.006$ & $ 3.7 \pm 0.2$ & $0.057 \pm 0.004 $ \\
	  $3.40  \pm 0.02 $ & $0.16  \pm 0.02 $ & $-1.5 \pm 0.2$ & $0.100 \pm 0.012 $ \\
	  $6.9  \pm 0.2 $ & $0.35 \pm 0.16$ & $0.5 \pm 0.2$ & ~ \\
      \tableline
    \end{tabular}
    \flushleft
    \tablecomments{Normalization is given in units of ct\per{s}\persq{cm}. Uncertainties are quoted at $68\%$ confidence. }
\end{table}
  
  The absorption line at $3.4\kev$ is the second most prominent line in
  spectrum. With a $\Delta\chi^2$ of $121$ and a simple significance ratio of
  $6.3\sigma$, this line is highly significant. The line again appears to be
  slightly asymmetric, with a shallower wing toward low energy and a steeper
  wing toward high energy. We repeated this analysis with a simpler
  blackbody plus power-law model, but again obtained the same significance
  and line asymmetry. 

  To verify if a $3.4\kev$ absorption line could be detected at any other time,
  we first stacked all data in which the burst emission dominated the continuum
  (i.e. ObsIDs 01 and 02, excluding the first pointing in which the line was
  already detected). The spectrum of this stacked dataset showed no evidence
  for absorption at $3.4\kev$, yielding a $95\%$ upper
  limit on the absolute line normalization of $|N|
  < 5\E{-5}$ct\per{s}\persq{cm}. Additionally, we also combined all
  observations in which the source spectrum had returned to its persistent
  shape (ObsIDs 05 through 12). Again, the energy spectrum showed no evidence
  of absorption to an upper limit of $|N|
  < 2\E{-5}$ct\per{s}\persq{cm}.
  
  To investigate if the absorption line showed any temporal evolution, we
  divided this observation into three equal duration sub-segments
  ($\approx160\s$ exposure each). The absorption line was clearly detected in
  each individual sub-segment with the same line profile in all spectra. Hence,
  no line evolution was observed. We also binned the data as a function of
  intensity by constructing an $8\s$ resolution light curve and sorting the
  light curve bins into three flux bins of ascending count-rate, each with equal
  exposure. Extracting spectra from the underlying event data associated with
  each flux bin, we again clearly resolve the absorption line in all spectra
  but do not find any evolution in the line profile.

  For the two remaining lines, we again use the simulation method to find that
  the emission line at $0.6\kev$ is significant ($\Delta\chi^2 = 27$), but that
  the line at $6.9\kev$ is not ($\Delta\chi^2 = 7.2$). The associated simple
  significance ratios for these two lines are $4.3\sigma$ and $1.9\sigma$,
  respectively. The detailed component parameters for these lines are reported
  in Table \ref{tab:spectral lines}.

\subsection{Emission line modeling}
\label{sec:pion}
\begin{figure}[t]
  \centering
  \includegraphics[width=\linewidth]{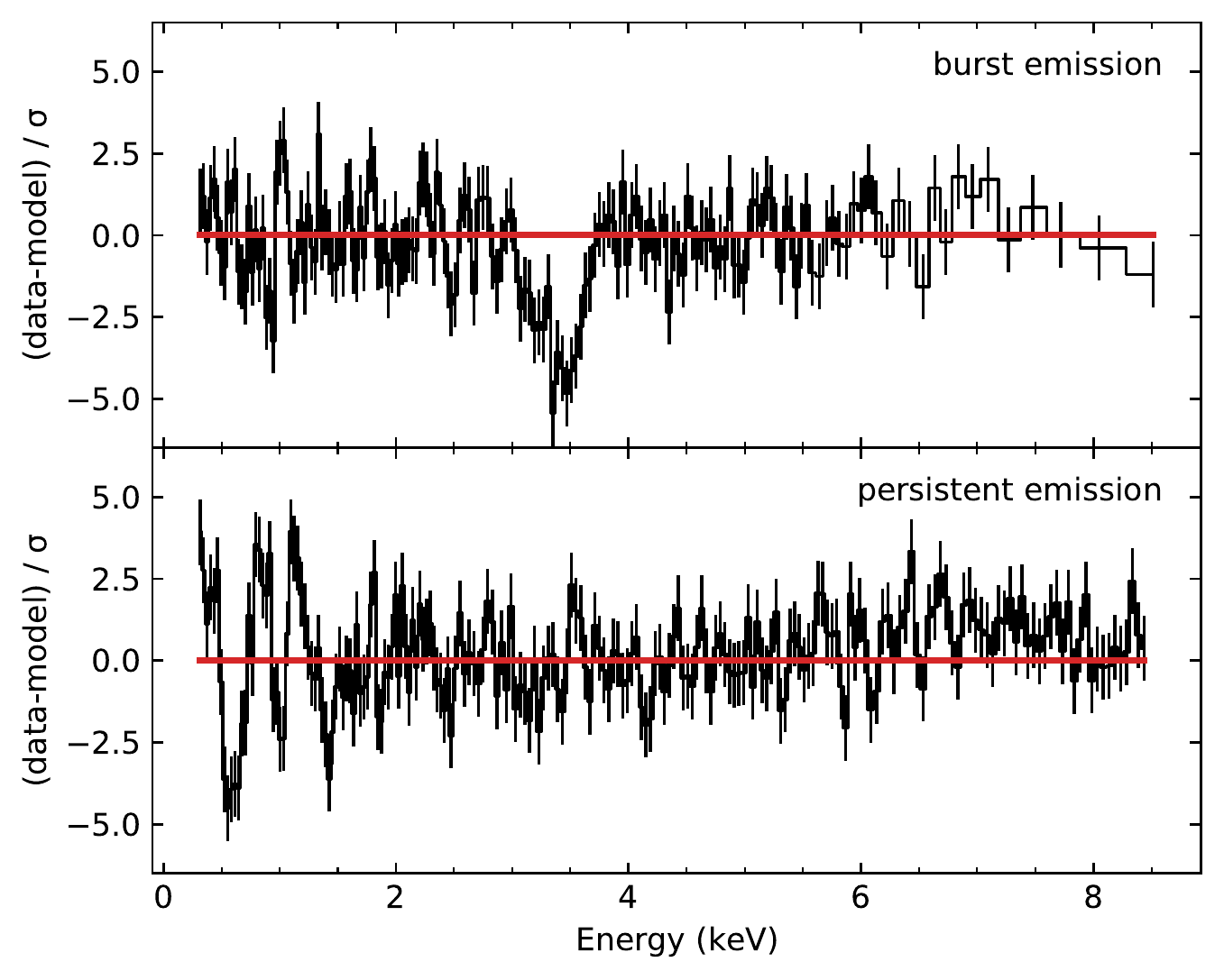}
  \caption{%
    Residuals of the best-fit \texttt{pionbb} model fits as applied to the
    burst emission spectrum (top) and the persistent emission spectrum (bottom).
  } 
  \label{fig:pion residuals}
\end{figure}

High resolution spectroscopy of the persistent emission has shown that the
$1\kev$ emission region consists of a continuum of narrow lines
\citep{Degenaar2017, Eijnden2018}. If a similar continuum of narrow lines is
responsible for the emission features observed in the burst spectrum, then that
may explain the apparent asymmetry of the $1\kev$ line. To investigate this
possibility, we attempted to fit the burst spectrum (Figure \ref{fig:spectral
lines}) using the photoionized plasma model \texttt{pion} \citep{Miller2015,
Mehdipour2016} which is available within \textsc{spex} \citep{Kaastra1996}.
Assuming a blackbody continuum illuminating a Compton-thin plasma, we
calculated \texttt{pion} model emission spectra on a parameter grid in
blackbody temperature, column depth, and ionization, which we then tabulated for
use in \textsc{xspec} (see \citet{Parker2019} for further details). In the
following, we will refer to this model as \texttt{pionbb}. When fitting to the
data, we adopted the reflection model for the continuum (as described in the previous section), and tied the blackbody temperature between the \texttt{pionbb} and continuum blackbody components. Because we did not calculate absorption features in our
model spectra, we further excluded the $3-4\kev$ region from the fit. This
model performed reasonably well, yielding a best-fit $\chi^2$ of 183.6 for 138
degrees of freedom for the parameters listed in Table \ref{tab:pion}.
As shown in the top panel of Figure \ref{fig:pion residuals}, the model
captured most of the emission around $1\kev$ and simultaneously accounted for
the $0.63\kev$ line as well.

We also attempted to fit the $1\kev$ emission feature observed in the
(non-burst) persistent emission with the \texttt{pionbb} model. We combined all
data collected after $t=4$, i.e. ObsID $05$ through $12$, and extracted
a spectrum between $0.3-9.0\kev$. We again modelled the continuum emission
using the reflection model (adopting the \texttt{relxillD} model as described
in Section \ref{sec:reflection fits}), but exchanged the Gaussian line
component for the \texttt{pionbb} model. This model did not yield an especially
good fit, resulting in a best-fit $\chi^2$ of $482$ for $201$ degrees of
freedom, mainly due to large residuals at low photon energies (see Figure
\ref{fig:pion residuals}, bottom panel). For comparison, adopting a simple
Gaussian line component resulted in best fit $\chi^2$ of 326. To ensure that
the poor fit did was not caused by the slightly different treatment of the disk
reflection component, we repeated the \texttt{pionbb} modelling using the
phenomenological continuum model as well. For both the burst and persistent
spectra we obtain the same outcome: the \texttt{pionbb} model works well for
the burst spectrum, but fails for persistent emission.

\begin{table}[t]
    \centering
    \caption{%
      Spectral parameters of the \texttt{pionbb} fits.
	\label{tab:pion}
    }
    \begin{tabular}{l l l l}
      \hline \hline
      Parameter & Value & Scale & Unit \\
      \tableline
      \sidehead{Burst spectrum}
      \tableline
      Ionization (log $\xi$) & $2.73_{-0.11}^{+0.07}$ & ~ & erg s \persq{cm} \\
      Column density (nH) & $1.19\pm0.76$ & $\times 10^{20} $ & \persq{cm} \\
      Temperate (kT) & $0.71 \pm 0.02$ & ~ & \kev \\
      Flux ($1-10\kev$) & $2.1\pm0.3$ & $\times10^{-11}$ & \fluxcgs \\
      \tableline
      \sidehead{Persistent spectrum}
      \tableline
      Ionization (log $\xi$) & $2.82\pm0.04$ & ~ & erg s \persq{cm} \\
      Column density (nH) & $1.0\pm0.7$ & $\times 10^{20} $ & \persq{cm} \\
      Temperate (kT) & $0.31 \pm 0.01$ & ~ & \kev \\
      Flux ($1-10\kev$) & $1.3\pm0.2$ & $\times10^{-12}$ & \fluxcgs \\
      \tableline
    \end{tabular}
    \flushleft
    \tablecomments{Best-fit parameters of the \texttt{pionbb} model as applied to the burst emission spectrum (Figure \ref{fig:spectral lines}) and the persistent emission spectrum (Figure \ref{fig:example spectra}). The associated residuals are shown in Figure \ref{fig:pion residuals}. See Section \ref{sec:pion} for more details. Uncertainties are quoted at $90\%$ confidence. }
\end{table}

\section{Discussion}
\label{sec:discussion}
We have presented a spectroscopic analysis of \src in the aftermath of one
of its energetic intermediate duration X-ray bursts. We detected a significant
absorption line in the source emission when the burst emission was still
bright. Further, we have investigated the spectral evolution as the source
moves through a dip in overall intensity directly after the burst has cooled.
In the following we discuss the implications of these findings.

\subsection{The intensity dip}
\begin{figure}[t]
    \centering
    \includegraphics[width=\linewidth]{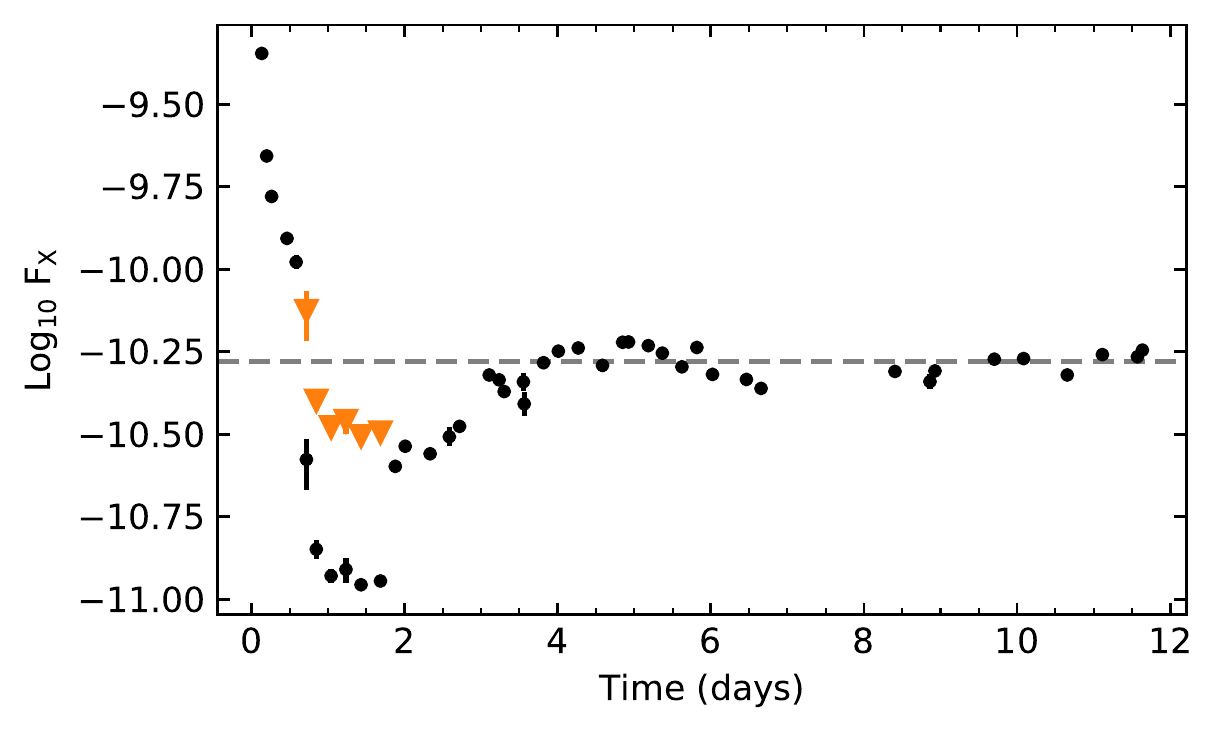}
    \caption{%
      X-ray flux of \src in the $1-10\kev$ band. The black points show the data
      as measured, with apparent discontinuous jumps at $t=0.6$ and $t=1.9$. The orange points show the six measurements between these two jumps when multiplied with an ad hoc constant factor of $2.8$.
    } 
    \label{fig:flux shift}
  \end{figure}

As the X-ray burst emission decays, we find that the source emission drops well
below the long-term persistent rate. This intensity dip lasts about three days,
starting at $t=0.6$\,d and ending at $t=3.5$\,d, and is further punctuated by two
sharp transitions in the intensity. These transitions are already apparent in
the light curve (Figure \ref{fig:lc}), where the count-rate drops sharply at
$t=0.6$\,d, and jumps up again at about $t=1.9$\,d. These same transitions are even
more pronounced in the evolution of the $1-10\kev$ X-ray flux (Figure
\ref{fig:pl spectra}), where the six pointings collected while the source
passes through its lowest intensities appear to be shifted in flux by
a constant factor. Indeed, if we increase these six flux measurements by an
(ad hoc) factor of $2.8$, then the resulting light curve becomes a smoothly
varying trend (see Figure \ref{fig:flux shift}). This suggests that there are
two superimposed parts to the intensity dip: firstly there is the wider three
day period during which the source drops below the persistent intensity, and
secondly there is the narrower $1.5$ day interval during which we are seeing
only about $35\%$ of the expected emission.

What could cause this complex evolution in the observed X-ray flux? The
evolution of the spectral parameters offer some clues here (Figure
\ref{fig:reflection spectra}). As the X-ray burst emission cools, the blackbody
temperature evolves smoothly to well below its normal persistent temperature.
Given that the blackbody is attributed to the neutron star \citep{Eijnden2018, Hernandez2019} 
its temperature is set by heating associated with the
impact of the accretion flow. Hence, our results immediately suggest that the
accretion onto the star must be inhibited. That the impact of an X-ray burst
could trigger a temporary change in the mass accretion rate is in itself not
surprising. There is ample observational evidence from both short and long
X-ray bursts that the accretion rate onto the star is enhanced while the X-ray
burst is brightly irradiating the disk \citep{Zand2013, Worpel2013,
Worpel2015}. 
There are a number of physical mechanisms through which the burst emission
could affect the mass flow through the disk \citep{Degenaar2018}: radiation
pressure of the burst emission could drive an outflow; the X-ray heating might
change the structure of the disk, likely increasing the scale height and
altering the density; and Poynting-Robertson drag could remove angular momentum
from the disk, allowing it to drain onto the neutron star. Recent numerical
simulations of the radiative burst-disk interaction indicate that the latter
two effects dominate the disk response to an X-ray burst \citep{Fragile2018,
Fragile2020}, causing the inner radius of the disk to move outward as the burst
rises to its peak intensity.
Following the bright phase of the burst, one might then expect that it takes
some time for the disk to replace the lost material, thus temporarily
suppressing the amount of matter that reaches the stellar surface. We suggest
that this disk-recovery process is causing the wider three day dip in the source
flux.

One obvious challenge to this interpretation is that three days is
a comparatively long time for a disk recovery process. The simulations of
\citet{Fragile2020} indicate the timescale at which the inner disk radius
recovers is similar to the decay timescale of the burst, which for \src is on
the order of hours, rather than days. However, they also find that the disk
structure remains perturbed even after the burst has decayed. Hence, the
observed $3$ day dip might not be associated with the recovery of the disk
inner radius, but rather with the relaxation of the disk structure to its
preburst equilibrium. It is important to keep in mind, however, that the
simulations of \citet{Fragile2020} investigate the disk response to a regular
X-ray burst, that is, its duration was on the order seconds. The intermediate
duration bursts of \src are much more energetic events,
so there is some uncertainty as to whether the results of
these simulations extrapolate: the longer X-ray bursts observed from \src
may induce a more pronounced disk response.

If this interpretation is correct, then we would expect that excess mass
accreted during the burst is roughly the same as the mass needed to restore the
disk. Equivalently, the energy budget of the observed flux reduction should
match the expected increase in the persistent flux during the burst. As
a consistency test, we therefore adopted the flux evolution with the $2.8$
correction factor applied, and measured the fluence over the $3$ day interval
of the intensity dip. This fluence is $8.6\E{-6}\fluecgs$, as compared to the
$1.25\E{-5}\fluecgs$ expected from a constant flux of $5.6\E{-11}\fluxcgs$.
Dividing the difference in fluence by the $10^3-10^4\s$ duration of the X-ray
bursts of \src \citep{Keek2017}, we find that the accretion emission must have
been inflated by a factor of $7-70$ during the X-ray bursts. While this range
is rather broad, it is consistent with the apparent increases in the persistent
emission of many bright PRE bursts \citep{Zand2013, Worpel2013}.

If the three day dip is indeed caused by a gradual restoration of the inner
accretion disk, then, one wonders why this effect is not more commonly
observed. Here the special characteristics of \src may play an important role
in making this recovery phase so pronounced. First, the amount of matter
evacuated from the inner disk through enhanced accretion will likely scale with
the intensity and duration of the X-ray bursts. As the intermediate duration
X-ray bursts of \src are among the most energetic helium bursts on record
\citep{Zand2019}, their impact on the disk will be more pronounced compared to
regular H/He bursts. 
Second, the ultra-compact nature of \src means its accretion disk is
comparatively small. Hence, it is plausible that the entire disk is affected by
the irradiating burst.
Third, in its persistent state, the accretion rate onto the stellar surface is
very low \citep{Keek2017}, thus, if a large amount of matter is depleted during
the bursts, the subsequent recovery is likely to have a longer timescale than
it would have in brighter sources. Fourth, \src is a millisecond pulsar that is
believed to have an active propeller mechanism \citep{Hernandez2019,
Bult2021b}. 
The radiation pressure and X-ray heating induced by the
burst will likely affect the same region of the disk where the magnetosphere 
couples to the accretion flow. The net effect this will have on the propeller
is difficult to predict, as competing effects might be expected. For instance,
the Poynting-Robertson drag removes angular momentum from the flow, which could
make it easier for matter to be ejected. On the other hard, if the disk is being
disrupted, that might affect the magnetic threading of the disk, potentially
reducing the efficiency with which the stellar magnetosphere can exert
a torque on the disk. Either way, an impact of the burst on the propeller
efficiency seems plausible and might explain why a burst from \src elicits
a different disk response as compared to other sources.

In interpreting the three day dip, we have assumed that the lowest intensity
observations were shifted in flux, but we have not yet addressed what could
cause this shift. 
The obvious interpretation would be that the evolving disk structure is
temporarily obscuring the central source. This scenario would fit with the
rapid transitions in flux, as well as the apparent constant flux scaling.
A problem with this interpretation, though, is that obscuration dips are
normally
associated with high-inclination sources. The inclination of \src, on the other
hand, is likely about $30-35\arcdeg$ \citep{Bult2021b}. More importantly, though,
the spectral evolution seen during the flux shift is not consistent with absorption. 

Considering the time-resolved spectral evolution, we see that the blackbody parameters evolve smoothly across both discontinuities
in flux. Hence, it would appear that the reprocessed emission is the main
driver for this effect. Indeed, none of the six pointings that make up the
flux-shifted epoch show evidence of a reflection component, or a $1\kev$
emission line. These non-detections could indicate a temporary change in the
inner disk geometry, e.g. a change in the disk scale height might cause
a change in disk anisotropy, or even shadow the outer parts of the disk from
the irradiating continuum emission. Considering the intensity of the disk
reflection, however, we see that when it is detected, the reflection component
consistently makes up about one third of the total X-ray flux. Hence, just
removing the reflection from the model cannot fully account for the observed
flux shift. The high photon energy tail must be affected also.
  
Another aspect to consider is that the character of the reflection spectrum
appears to be different before and after the flux shift: the early data prefers
the blackbody reflection model, whereas the later data prefers the power law
reflection model. Simultaneously, the early data has a lower power law photon
index than the late data. This suggests that the geometry of the Comptonization
medium is different while the burst is bright as compared to the persistent
state. Possibly, then, the period of the flux shift is related to the changing
corona. 

The evolution of a corona under the influence of an X-ray burst is shaped by
several competing effects \citep[see][for a recent
investigation]{Speicher2020}. The flood of soft photons from the burst should
effectively cool the corona. As the geometry of the corona changes, however, it
will intercept a different fraction of the burst emission, modifying the
cooling rate. Further, the enhanced disk accretion rate may provide
a temporary boost to the coronal heating rate, partially counteracting the
radiative cooling.

With these interaction mechanisms in mind, we can speculate how the changing corona might lead
to the observed spectral evolution. While the burst is bright, it likely causes
the corona to condense into a high density state. This would be consistent with
the reduced power-law emission and its lower photon index. As the burst cools
to near the persistent intensity, however, it can no longer hold the disk and
corona in their perturbed states. As the disk begins to recover, the corona
might likewise inflate back into a more extended geometry. Due to the resulting
decrease in density, the corona may be efficiently cooled by the latent burst emission, with few hard photons
visible to either the disk or the observer. Once the disk has sufficiently
relaxed, the density of the corona recovers, allowing for the (re-)emergence of
the power-law continuum and the associated disk reflection.

\subsection{Emission lines}
We observed two spectral emission lines in the burst spectra of \src. The most
prominent of the two is the $1\kev$ emission line, which was also observed
in the previous bursts, as reported by \citet{Degenaar2013}
and \citet{Keek2017}. These
authors interpreted the line as a feature of the reflection spectrum, possibly
associated with Fe-L band transitions or perhaps with Ne\,\textsc{x}. A similar
$1\kev$ emission line has also been observed in bright PRE bursts of 4U
1820$-$30 \citep{Strohmayer2019} and SAX J1808.4$-$3658 \citep{Bult2019b}. As
in these previous works, we find that reflection models are not able to account
for the observed intensity in the $1\kev$ line. This does not rule out
a reflection origin, however, as the line strengths depend greatly on the
assumed composition of the disk, where even state-of-art reflection models are
still imperfect \citep{Ballantyne2004, Madej2014, Keek2017}. 

A $1\kev$ emission line is also present in the persistent emission of \src,
however, high-resolution spectroscopy has revealed that this feature is
a continuum of several narrow lines \citep{Degenaar2017, Eijnden2018} rather
than a single broad line. While a reflection origin remains plausible,
\citet{Eijnden2018} suggest that these lines may also be produced by an outflow
or through shocks in the accretion column.  Presumably, the physical mechanism
behind the $1\kev$ emission is the same for both the burst and the persistent
emission, so we can speculate that the line we measure in this work is also
a superposition of several more narrow lines. 

We also detected a weaker emission line at $0.63\kev$ in the X-ray burst spectrum. A similar feature has not previously been reported in either the
2012 or 2015 bursts \citep{Degenaar2013, Keek2017}. \citet{Eijnden2018} do
report excess emission around these energies in the persistent emission, and
attribute this to an enhanced oxygen abundance in the disk. Line emission
around $0.6-0.7\kev$ has also been observed in other ultra-compact X-ray
binaries, notably 4U 0614+091 \citep{Madej2010}, 4U 1543--624 \citep{Madej2011,
Ludlam2019}, although those examples are much more prominent than what we report here. In both cases this line is interpreted as O\,\textsc{viii}
emission associated with the disk reflection spectrum. 

In an attempt to quantity the line emission with a physical model, we fit these
features with a \texttt{pion} model for the line emission of a photoionized
plasma illuminated by a blackbody continuum. This model was able to
simultaneously describe both the $0.63\kev$ and $1\kev$ emission, with the
latter being due to a continuum of narrow line features (as in the persistent
emission). The plausible interpretation of this model is that the lines
originate in an outflow that is illuminated by the burst emission, although it
remains possible that the \texttt{pion} model is simply compensating for some
of the deficiencies in the disk reflection models. Either way, the success of
this model would suggest that both emission lines share the same origin.
Applying the \texttt{pion} model to the persistent (non-burst) emission
spectrum, however, we found that it failed to adequately describe the broader
$1\kev$ feature observed there. Specifically, the best
fit to the persistent emission spectrum leaves structural residuals at low
photon energies (Figure \ref{fig:pion residuals}), indicating that the model is
not quite able to produce a line as broad as seen in the data.  We could
attribute this change in model performance to the changing continuum emission,
although it may also indicate a structural change in emitting plasma. A more
detailed comparison of the emission lines in burst and persistent spectra may
shed light on the precise origin or these lines, however such a study is beyond
the scope of this work. 

\subsection{Absorption line}
  We detected a highly significant absorption line at $3.4\kev$ in the first
  $472\s$ duration pointing of \src. The absorption line is observed while the
  thermal burst emission dominates the spectrum: at $3.4\kev$ the blackbody
  component accounts for approximately $85\%$ of the continuum emission.
  Conversely, the depth of the absorption line has a minimum at about $80\%$ of
  the continuum, exceeding the intensity in the non-burst emission by a factor
  of $1.5$. Hence, this line cannot be associated with reprocessing in the disk
  or corona. 
  Two plausible origins for this line remain: it could be associated
  with absorption in a burst induced outflow, or it could originate from heavy
  elements on the stellar surface. Either way, it is likely to be a signature
  of burning ashes produced in the burst.

\begin{table}[t]
    \centering
    \caption{%
	Hydrogen-like K$\alpha$ transition energies.
	\label{tab:kalpha}
    }
    \begin{tabular}{l l l}
      \hline \hline
	  Element & Energy (keV) & 1+z  \\
      \tableline
      S  & 2.62 & 0.77 \\
      Ar & 3.32 & 0.98 \\
      K  & 3.70 & 1.09 \\
      Ca & 4.11 & 1.21 \\
      Sc & 4.53 & 1.33 \\
      Ti & 4.97 & 1.46 \\
      \tableline
    \end{tabular}
    \flushleft
    \tablecomments{The right most column lists the gravitational redshift (1+z)
    for that element to produce the observed $3.40\pm0.02\kev$ absorption
    line.}
\end{table}

  Spectral features in the burst emission have long been sought after, because,
  in principle, they offer a direct probe of the stellar compactness. A line
  originating from the stellar surface will be gravitationally redshifted
  by a factor $1+z = E_0/E$, which depends on stellar mass and radius as
  \begin{equation}
    \frac{M}{R} = \frac{c^2}{2G} (1 - (E/E_0)^2),
  \end{equation}
  with $R$ and $M$ the stellar mass and radius, $E$ the observed line energy,
  and $E_0$ the rest-frame line energy. In practice, however, finding and
  identifying such features has proven challenging. Early burst studies with
  Tenma and EXOSAT reported absorption lines at $4.1\kev$ \citep{Waki1984,
  Nakamura1988, Magnier1989}, and at $5.7\kev$ \citep{Waki1984}, however, more
  sensitive instrumentation has never been able to confirm those detections.
  \citet{Cottam2002} reported on a set of narrow absorption lines in the
  stacked XMM-Newton spectra of EXO $0748-676$. These lines, however, were
  later found to be incompatible with a surface origin \citep{Lin2010}. Using
  NuSTAR, \citet{Barriere2015} found a narrow $5.5\kev$ absorption line in GRS
  1741.9$-$2853, although at $1.7\sigma$ its significance is marginal. NICER
  observations of 4U 1820$-$30 revealed narrow absorption lines at $1.7\kev$
  and $3.0\kev$ \citep{Strohmayer2019} which were attributed not to the stellar
  surface, but to a PRE driven wind.  

  The detection of absorption edges in burst spectra are more robust
  \citep{Zand2010, Kajava2017}, and are attributed to heavy metals dredged up
  during the PRE phase of the X-ray burst.  
  During the thermonuclear runaway powering the X-ray burst, the nuclear
  burning chains will seed the stellar envelope with burning ashes that
  primarily consist of heavy elements with atomic numbers in the $30\sim60$
  range \citep{Schatz2001, Brown2002, Woosley2004}. Convective mixing raises
  these elements to sufficiently shallow depths that they may be ejected in the
  super Eddington wind generated by the PRE phase of the burst
  \citep{Weinberg2006, Yu2018}. Applying this interpretation to the $3.4\kev$
  line observed in \src, we find a possible identification might be hydrogen-like
  Ar\,\textsc{xviii} which has a line energy at $3.32\kev$ (see Table
  \ref{tab:kalpha}). The modest blueshift could be explained if the absorption
  line is created at a large height above the stellar surface, such that the
  line shift is mainly a function of the outflow velocity. 

  The wind interpretation has a number of issues, the first of which is
  the timing of the absorption line detection. The line is observed only in the first
  observation, which was collected about three hours after the MAXI/GSC trigger. 
  Although we did not observe the 2020 burst from \src during its peak intensity,
  the large energy budget of these bursts makes it reasonable to assume that all
  bursts show photospheric radius expansion, and probably even super-expansion
  \citep{Degenaar2013, Keek2017}. For the 2015 burst, \citet{Keek2017} estimated
  that the PRE phase lasted about $200\s$. Hence, if the 2020 burst followed
  a comparable evolution, then the PRE phase and its associated wind would have
  ended well before the first \nicer observations were collected. 
  A second issue is that the line is relatively broad. If the absorption line
  originates in a wind, then the width of the line should be dynamic,
  suggesting a material velocity of $0.11c$. In contrast, the expected wind
  velocity is only $\lesssim0.01c$ \citep{Yu2018}. Finally, if the
  absorption line has a dynamic origin, then one might expect an evolution in
  the line profile with time or flux, as was reported by \citet{Strohmayer2019}
  in the case of 4U 1820$-$30, for instance. The $3.4\kev$ absorption line does not show any such
  evolution.

  Alternatively, we might associate the absorption line with burning ashes that
  remain on the stellar surface. Modeling suggests that the heavy elements
  created during the burst may linger on the surface \citep{Yu2018}, where they
  can imprint on the observed burst emission even after the PRE phase has ended.
  Given the expected redshift factor of $1+z\approx1.3$ for a canonical neutron
  star, the line could be associated with the $K\alpha$ transition of
  hydrogen-like
Sc\,\textsc{xxi} (see Table \ref{tab:kalpha}). A problem with this
identification, however, is that \src is an ultra-compact binary
\citep{Strohmayer2018a}, hence the accreting material should be depleted of
hydrogen. One would therefore expect the nuclear reaction chain powering the
X-ray burst to proceed through a series of alpha captures, which flows through
$\prescript{36}{}{\rm Ar}$, $\prescript{40}{}{\rm Ca}$, and $\prescript{44}{}{\rm
Ti}$, and does not produce Sc. Possibly, the nuclear reactions could seed the
envelope with a small amount of protons, which may then act as catalysts for
the production of non-alpha chain nuclei \citep{Weinberg2006}. Even in this
case, however, the various isotopes of Sc are not expected to be 
produced in abundance \citep{Parikh2008, Cyburt2016}. Instead, $\prescript{40}{}{\rm Ca}$ or
$\prescript{44}{}{\rm Ti}$ are likely the dominant elements in the burning
ashes of a helium burst \citep{Weinberg2006, Yu2018}, making the Ca\,\textsc{xx} or Ti\,\textsc{xxii}
line transitions the more probable identifications for the observed absorption
line (see Table \ref{tab:kalpha}).
Assuming a $10-15$ km neutron star radius, a line
identification with these two elements would imply a stellar mass of
$1.07-1.61\msol$, or $1.80-2.7\msol$ respectively. 

  General relativistic effects will not only introduce a redshift, but will
  also shape the observed line profile, introducing a dependence on the stellar
  spin, the oblateness, and the opening angle between the observer's line of sight
  and the stellar rotation axis \citep{Ozel2003}. Together, these effects will cause
  the observed line to broaden and skew toward an asymmetric profile that
  favors higher photon energies. Qualitatively, these effects are consistent with
  the observed $3.4\kev$ absorption line, which has a broadened profile and appears
  to be slightly skewed toward higher energies. Quantitatively, however, the width
  of the measured line converts to a full width at half maximum of $0.38\kev$. Comparing
  this width to line profiles obtained through ray-tracing simulations
  \citep{Baubock2013, Nattila2018}, the observed line appears to be much
  broader than expected for a $164$ Hz millisecond pulsar.

The width of the absorption line might be reconcilable with both a PRE wind and
stellar surface origin if we are not actually observing a single line, but
rather a superposition of two or more unresolved lines. The energy resolution
of \nicer at $3.4\kev$ is about $100$ eV, so it is feasible that the absorption
feature really consists of two or more closely spaced narrow lines that are
blended together. Given that the absorption line was observed late in the
cooling tail of the burst, the wind interpretation remains problematic, so we
favor the surface origin. A number of questions remain, though. While the
modelling of burst reaction chains suggests that either $\prescript{40}{}{\rm
Ca}$ or $\prescript{44}{}{\rm Ti}$ may be present in the neutron star envelope
with high abundances \citep{Weinberg2006, Yu2018}, it is not clear how long
such elements might linger near the stellar surface. Further, we caution that
these works are based on ignition depths that go down to $5\E{9}$ g\persq{cm}.
The bursts of \src have an ignition depth that is an order of magnitude larger
\citep{Keek2017}, introducing some uncertainty as to whether the simulation
results extrapolate to these events.  Hence, a more targeted theoretical
investigation will be needed to identify the origin of these lines with
confidence. 
  
\section{Conclusions}
We have presented a spectroscopic analysis of \nicer observations in the
aftermath of a bright intermediate duration X-ray burst from \src. 
These observations revealed that the flux of \src was severely depressed
over the three day period following the X-ray burst. We interpreted this
flux evolution as an effect of the accretion disk and corona gradually relaxing
back to their persistent equilibrium states, after being perturbed by the
energetic X-ray burst.

We further detected a weak emission line at $0.64\kev$ and a strong emission
line at $1\kev$ in the X-ray burst spectrum. We found that a photoionized
plasma illuminated by a blackbody continuum is able to simultaneously describe
both lines, suggesting they share a common origin. The precise nature of the
line emitting plasma remains unclear at this time, however, and could be
attributed to an ionized outflow or disk reflection.

Finally, we detected a prominent $6.4\sigma$ absorption line centered at
$3.4\kev$. We tentatively attribute this line to burning ashes on the stellar
surface, possibly due to $\prescript{40}{}{\rm Ca}$ or $\prescript{44}{}{\rm
Ti}$.

\clearpage

\begin{acknowledgments}
\begin{nolinenumbers}
This work was supported by NASA through the \nicer mission and the
Astrophysics Explorers Program, and made use of data and software 
provided by the High Energy Astrophysics Science Archive Research Center 
(HEASARC). PB further acknowledges support from the NICER Guest Observer program
(80NSSC21K0128) and the CRESST II cooperative agreement (80GSFC21M0002).
DA acknowledges support from the Royal Society. 
\end{nolinenumbers}
\end{acknowledgments}

\facilities{ADS, HEASARC, NICER}
\software{heasoft (v6.27.2), nicerdas (v7a)}

\bibliographystyle{fancyapj}

\end{document}